\documentclass[a4paper]{EslabStyle}
\usepackage{epsfig}
\def\be{\begin{equation}}
\def\ee{\end{equation}}

\def\etal{{\it et al.~}}

\def\HH{${\rm {H_2}}\,\,$}

\def\etal{{\it et al.~\/}}

\def\ie{{\it i.e.~\/}}

\def\ltsima{$\; \buildrel < \over \sim \;$}
\def\simlt{\lower.5ex\hbox{\ltsima}}
\def\gtsima{$\; \buildrel > \over \sim \;$}
\def\simgt{\lower.5ex\hbox{\gtsima}}

\begin{document}

\title{Feedback Processes in the Early Universe}

\author{A. Ferrara\inst{1}, B. Ciardi\inst{2}, S. Marri\inst{3}  \and P. Todini 
  \inst{2}} \institute{Osservatorio Astrofisico di Arcetri, Firenze, Italy
  \and Max-Planck-Institut f\"ur Astrophysik, Garching, Germany 
  \and Dipartimento di Astronomia, Universit\'a di Firenze, Italy}       

\maketitle 

\begin{abstract}
Feedback effects due to massive stars and supernovae in the first objects 
are shown to strongly 
regulate both galaxy formation/evolution and the reionization process. Here
we review the most important ones in some detail.
We discuss how Type II supernovae can be used as tracers of the first objects
and detected with NGST, for which we predict supernova number counts including
the effects of gravitational lensing. Preliminary results
on the formation of dust in the ejecta of supernovae of primordial composition
are also presented. 
We finally turn to the consideration of the process of inhomogeneous reionization
due to primordial stellar sources
by means of high resolution numerical simulations, allowing for a
self-consistent treatment of the above feedback processes.
These simulations allow us to draw conclusions on the evolution and 
epoch of reionization and about the fate of reionizing objects. We
conclude that a large fraction ($\approx 99$\%) of collapsed objects must
be dark at redshift around eight. 
\end{abstract}

\section{The First Objects}
  
As the temperature of the cosmic bath decreases, atoms start to recombine
and therefore decouple from  CMB radiation at redshift $\approx 1100$.
The baryonic Jeans mass after this event is given by (assuming $\Omega =1$)
\begin{equation}
\label{mj}
M_j \simeq 6\times 10^4 \left({1+z\over 30}\right)^{-3/2} \left({T\over 500 {\rm
K}}\right)^{3/2} \Omega_b M_\odot,
\end{equation}
where $T$ is the gas temperature and $\Omega_b$ the baryon density parameter.
Masses larger than $M_j$ are gravitationally unstable and should, in principle,
collapse. However, in order for the actual collapse to occur a more severe
condition must be satisfied, \ie that the cooling time of the gas is shorter than the
Hubble time
at that epoch. In fact, radiative losses provide the only way for the gas to lose
pressure and to settle down in the potential well of the dark matter halo.
Since the virial temperature corresponding to the masses of the first objects
(PopIII) is typically
$\simlt 8000$~K, cooling by hydrogen Ly$\alpha$ excitation is strongly
quenched,  and the only viable coolant in a primordial H-He plasma is molecular
hydrogen.
\HH is produced during the recombination phase, but its relic
abundance is very small.
Primordial H$_2$ forms with a fractional abundance of $\approx 10^{-7}$
at redshifts $\simgt 400$ via the H$_2^+$ formation channel. At redshifts
$\simlt 110$, when the Cosmic Microwave Background radiation (CMB) intensity
becomes weak enough to allow for significant formation of H$^-$ ions, a
primordial fraction of $f_{H_{2}}\approx 2\times 10^{-6}$ (Galli \& Palla 1998)
is produced for model universes that satisfy the
standard primordial nucleosynthesis constrain $\Omega_b h^2 = 0.0125$
(Copi, Schramm \& Turner 1995), where $\Omega_b$ is the baryon density parameter
and $H_0= 100 h$~km~s$^{-1}$~Mpc$^{-1}$ is the Hubble constant.
This primordial fraction is usually lower than the one required for the
formation of Pop~III objects, but during the collapse phase 
the molecular hydrogen content can reach high enough values to
trigger star formation.
On the other hand, objects with virial temperatures (or masses) above
that required for the hydrogen Ly$\alpha$
line cooling to be efficient, do not rely on  H$_2$
cooling to ignite internal star formation.
Thus, the the fate of a
virialized lump depends crucially on its ability to rapidly increase
its \HH content during the collapse phase.
Tegmark \etal (1997) have addressed this question in great detail by
calculating the evolution of the \HH abundance for different halo masses
and initial conditions for a standard CDM cosmology.
They conclude that if the prevailing conditions are such
that a molecular hydrogen fraction of order of $f_{H_2}\approx
5 \times 10^{-4}$ is produced during the collapse, then the lump will cool, fragment and eventually
form stars. This criterion is met only by larger halos implying that for each
virialization redshift there will exist
some critical mass, $M_{crit}$, such that protogalaxies with total mass
$M>M_{crit}$ will be able to form stars
and those with $M<M_{crit}$ will fail. As an example, a 3$\sigma$
fluctuation of a Cold Dark Matter primordial spectrum, has $M_{crit}\approx 10^6 
M_\odot$ and collapses at $z\approx 30$.

The first objects affect the subsequent galaxy formation, reionization and
metal enrichment of the universe through their feedback effects. Generally
speaking, two types of feedback, {\it radiative} and {\it stellar}, can be at work.
They are related to the ionizing/ \HH--photodissociating radiation and mechanical
energy input from massive stars and supernovae, respectively. We discuss them
in detail in the following. 

\section{Radiative Feedback}

As stars are formed inside the first collapsed objects,  their ionizing
radiation produces HII regions in the surrounding IGM, thus triggering the
reionization of the universe. In addition, their UV photons start to
photodissociate \HH   molecules in the neighbor objects. We have seen that 
\HH is a crucial species for the cooling to occur, and the lack of it might
prevent the collapse of small protogalaxies. This  process is known as {\it radiative 
feedback}.
Particularly relevant is       the soft UV background in the LW bands,
as by
dissociating the H$_2$, it could influence the star formation
history of other small objects preventing their cooling.

Ciardi, Ferrara \& Abel (2000) have shown that the UV flux from these objects
results in a soft (Lyman -- Werner band) UV background (SUVB), $J_{LW}$, whose
intensity
(and hence radiative feedback efficiency) depends on redshift.
At high redshift the radiative feedback can be induced also by the direct
dissociating flux from a nearby object.
In practice, two different situations can occur: i) the collapsing
object is outside the dissociated spheres produced by preexistent objects:
then its formation could be affected only by the
SUVB ($J_{LW,b}$), as by construction the direct flux
($J_{LW,d}$) can only dissociate molecular hydrogen on time
scales shorter than the Hubble time inside this region;
ii) the collapsing object is located
inside the dissociation sphere of a previously collapsed object:
the actual dissociating flux in this case is essentially given by
$J_{LW,max}=(J_{LW,b}+J_{LW,d})$.
It is thus assumed that, given a forming Pop~III, if the incident
dissociating flux ($J_{LW,b}$ in the former case, $J_{LW,max}$ in the
latter) is higher than the minimum flux required for negative
feedback ($J_{s}$), the collapse of the object
is halted. This implies the existence of a population of "dark objects"
which were not able to produce stars and, hence, light.

To assess the minimum flux required at each redshift to drive the
radiative feedback Ciardi \etal (2000, CFGJ) have performed non-equilibrium
multifrequency radiative transfer calculations for a stellar
spectrum (assuming a metallicity $Z=10^{-4}$) impinging onto a homogeneous gas
layer, and studied the evolution of the following nine species:
H, H$^-$,
H$^+$, He, He$^+$, He$^{++}$, H$_2$, H$_2^+$ and free electrons for
a free fall time.
We can then define
the minimum total mass required for an object to self-shield from
an external flux of  intensity $J_{s,0}$ at the Lyman limit, as
$M_{sh}=(4/3) \, \pi \langle\rho_{h}\rangle R_{sh}^3$, where
$\langle\rho_{h}\rangle=\langle \rho_b\rangle \Omega_b^{-1}$ is
the mean dark matter density of the halo in which the gas collapses;
$R_{sh}$ is the shielding radius beyond which molecular hydrogen
is not photodissociated and allowing the collapse to take place
on a free-fall time scale.

\begin{figure}[ht]
  \begin{center}
    \epsfig{file=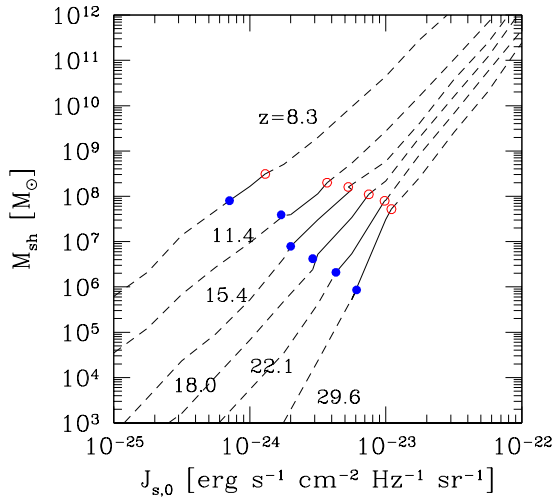, width=9cm}
  \end{center}
\caption{\label{fig1}{Minimum total mass for
self-shielding from an external
incident flux with intensity $J_{s,0}$ at the Lyman limit. The curves
are for different redshift: from the top to the bottom $z=$8.3, 11.4,
15.4, 18.0, 22.1, 29.6. Circles show the value of $M_H$ (open) and
$M_{crit}$ (filled). Radiative feedback works only in the solid portions
of the curves. }}
\end{figure}

Values of $M_{sh}$ for different values of $J_{s,0}$ have
been obtained at  various redshifts (see Fig. \ref{fig1}) and will be then used in
the reionization calculations discussed below.
Protogalaxies with masses above $M_H$ for    which
cooling is predominantly contributed by Ly$\alpha$ line cooling
will not be affected by the negative feedback studied here.
The collapse of very small objects with mass $<
M_{crit}$ is on the other hand made impossible by the
cooling time being longer than the Hubble time.
Thus the only mass range in which
negative feedback is important lies approximately
in $10^6-10^8 M_\odot$, depending on redshift. In order for the negative
feedback to be effective,
fluxes of the order of $10^{-24}-10^{-23}$erg s$^{-1}$ cm$^{-2}$
Hz$^{-1}$ sr$^{-1}$ are required.
These fluxes are typically produced by a Pop~III with baryonic mass $10^5
M_{b,5} M_\odot$
at distances closer than $\simeq 21-7\times M_{b,5}^{1/2}$~kpc for the
two above flux values, respectively, while
the SUVB can reach an intensity in the above range only after $z \approx 15$.
This suggests that at high $z$ negative feedback is driven primarily by the
direct irradiation from neighbor objects in regions of intense clustering,
while only for $z \le 15$ the SUVB becomes dominant.

\section{Stellar Feedback}

Once the first stars
have formed in the host protogalaxy, they can deeply influence the
subsequent star formation process through the effects of mass and energy
deposition
due to winds and supernova explosions.
While low mass objects may experience a blowaway, expelling their entire gas
content
into the IGM and  quenching star formation, larger objects may instead be able
to at least partially retain their baryons.
However, even in this case the blowout induces a decrease of the star
formation rate due to the global heating and loss of the galactic ISM.
These two regimes are separated by a critical mass, $M_{by}$ (CFGJ, Ferrara
\& Tolstoy 2000, MacLow \& Ferrara 1999). 
For the relatively small objects present
during the reionization epoch $30\simgt z \simgt 10$ the importance of these
{\it stellar feedbacks} can hardly be overlooked. To understand the role of stellar
feedback,
let us consider a collapsed object with mass lower than $M_{by}$.
Then the star formation is suddenly halted as the
entire gas content is removed. In this
case, the ionizing photon production will last only for a time interval of order
$t_{OB} \approx 10^7$ yr, the mean lifetime of the massive stars produced
initially.
After this phase, the ionized gas around the source will start to recombine at a
fast rate as a result of the highly efficient high $z$ Compton cooling,
rapidly decreasing the temperature inside the ionized region. Due to the
short lifetime and  recombination time scales, these object will only
produce transient HII regions which will rapidly disappear.
In fact, the recombination time scale, when the Compton cooling
is taken into account, is of order $t_{rec} \approx (1-50) \times 10^6$~yr
at redshift $z$$\approx$30-10, respectively,
and therefore much shorter than the corresponding Hubble time.
The dissociating photon production will nevertheless continue
for a longer time, due to the important contribution of long-lived
intermediate mass stars formed in the same initial star
formation burst. This, combined with the fact that there is no
efficient mechanism available to re-form  the destroyed  H$_2$ in the IGM
analogous to H recombination, implies that the dissociation is not
impeded by blowaway and the contribution from these small
objects should be yet accounted for.
However, this is not necessary. In fact, after blowaway,
H$_2$ is efficiently formed in the
shocked IGM gas, cooling under non equilibrium conditions (Ferrara 1998).
The final radius of the cooled shell behind which molecular hydrogen  is formed
by this process is:
\begin{equation}
R_s \simeq 224  M_{b,5}^{1/5} (1+z)^{-19/10} \; {\rm kpc}.
\label{rs}
\end{equation}
We notice that the formation process produces an amount of
\HH roughly similar to the one destroyed by radiation (Ferrara 1998).
A lower limit to the amount of molecular hydrogen   produced in an
explosion is readily found to be equal to $M^+_{H_2}(z)\simeq
(4\pi/3) \rho_b R_s^3 (2f) $ or
\begin{eqnarray}
\label{mh2+}
M^+_{H_2}(z)\simeq
102 f_{b,8}^{3/5}\Omega_{b,5}(1+z)_{30}^{-27/10} M_6^{1/5}
f_6 h^{7/10} {\rm M}_\odot.
\end{eqnarray}
The UV/ionizing radiation from Pop~III
massive stars previous to blow-away will produce both a HII region and a region
of photodissociated intergalactic \HH   (radius $R_d$) in which
the object is embedded. The radius $R_d$ can be defined by
requiring  that the photodissociation timescale $t_d$
is shorter than $t_H$. This condition yields the definition
$R_d = S_{LW}^{1/2} (1+z)^{-3/4} h^{-1/2}$, where $S_{LW}=
\beta S_i(0)$ is the UV photon flux in the \HH   Lyman-Werner (LW) bands
(11.2--13.6~eV),
assumed here to be proportional to the flux of Lyc photons, $S_i(0)$,
just before the massive stars explode. 
The value of the constant $\beta$ depends somewhat on the IMF
and on the evolutionary stage of the stellar cluster, but its
value should be reasonably close to unity.
We estimate $S_i(0)$ to be
\begin{equation}
\label{S}
S_i(0)\simeq 10^{47} f_{uvpp,48} f_{esc,20} \Omega_{b,5} f_{b,8} M_5
{\rm ~s}^{-1},
\end{equation}
where $f_{uvpp} = 48000 f_{uvpp,48}$ is the UV photon production
per collapsed proton efficiency (Tegmark \etal 1994) and $f_{esc}=
0.2 f_{esc,20}$ is the escape fraction of such photons.
It is worth noticing that the latter quantity is rather uncertain. 
Recently, Dove, Shull \& Ferrara (2000) have estimated the fraction of ionizing
photons emitted by OB
associations that escapes the H~I disk of our Galaxy into the halo and
intergalactic medium (IGM) by solving the time-dependent radiation transfer
problem of stellar
radiation through evolving superbubbles within a smoothly varying H~I
distribution. They find that the
shells of the expanding superbubbles quickly trap or attenuate the ionizing
flux, so that most of the escaping radiation escapes shortly after the
formation of the superbubble. Superbubbles of large associations can blow
out of the H~I disk and form dynamic chimneys, which allow the ionizing
radiation to escape the H~I disk directly. However, blowout occurs when the
ionizing photon luminosity has dropped well below the association's maximum
luminosity.  For a coeval star-formation history, the total fraction of
Lyman Continuum photons that escape both sides of the disk in the solar
vicinity is $f_{esc}\approx 0.15 \pm 0.05$; for a Gaussian star formation
history, $f_{esc}\approx 0.06 \pm 0.03$. 

\begin{figure}[ht]
  \begin{center}
    \epsfig{file=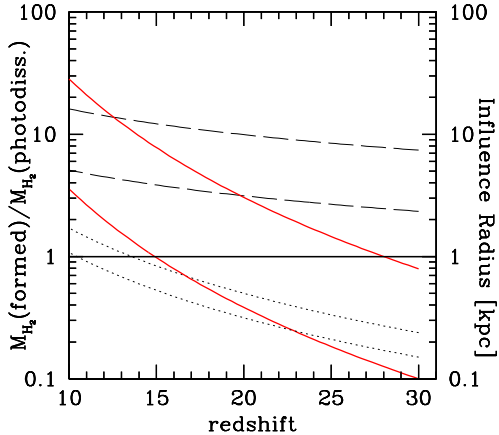, width=10cm}
  \end{center}
\caption{\label{fig2}{Ratio between the \HH   mass formed
and destroyed by a Pop~III object as a function of the
SNII explosion redshift (solid lines); values larger
than one for the ratio define the epochs where Pop~IIIs
have a {\it positive feedback} on galaxy formation. Also shown are
the shell (proper) radius at cooling, $R_s$ (dotted), and the
photodissociation (proper) radius, $R_d$ (dashed). The upper set of curves
refers to objects of mass $M=10^6 M_\odot$, whereas the bottom one corresponds
to larger objects, $M=10^7 M_\odot$.}}
\end{figure}

With these assumptions, it follows that
\begin{equation}
\label{rd}
R_d \simeq (\beta S_{47})^{1/2} (1+z)_{30}^{-3/4} h^{-1/2}
{\rm ~kpc},
\end{equation}
where $S_{47}= S_i(0)/10^{47} {\rm s}^{-1}$.
Thus, the ratio between the \HH   mass produced and destroyed
by Pop~IIIs is
\begin{equation}
\label{m+m-}
{M^+_{H_2}(z)\over M^-_{H_2}(z)}= \left({f\over f_{IGM}}\right)
\left[{R_s\over R_d}\right]^3.
\end{equation}
The previous relation is graphically displayed in Fig. \ref{fig2}
along with the values of $R_s$ and $R_d$. From that plot
we see that objects of total mass $M_5=10$ produce more \HH  
than they destroy for $z\simlt 25$; larger objects ($M_5=100$)
provide a similar positive feedback only for $z\simlt 15$,
since they are characterized by a higher $R_d/R_s$ ratio.
However, since in a hierarchical model larger masses form
later, even for these objects the overall effect should be
a net \HH   production.

As inside
$R_s$  \HH is re-formed, the net effect on the destruction of
intergalactic \HH of blown away objects is negligible (if not positive)
and we can safely neglect them in the subsequent calculations.

In conclusion, low mass objects with mass $< M_{by}$
produce ionization regions which last only for a recombination time;
larger objects can survive but their
star formation ability is impaired by ISM loss/heating. 
The detailed derivation of $M_{by}$ can be found in CFGJ.

\section{Evolutionary Tracks due to Feedbacks}

\begin{figure}[ht]
  \begin{center}
    \epsfig{file=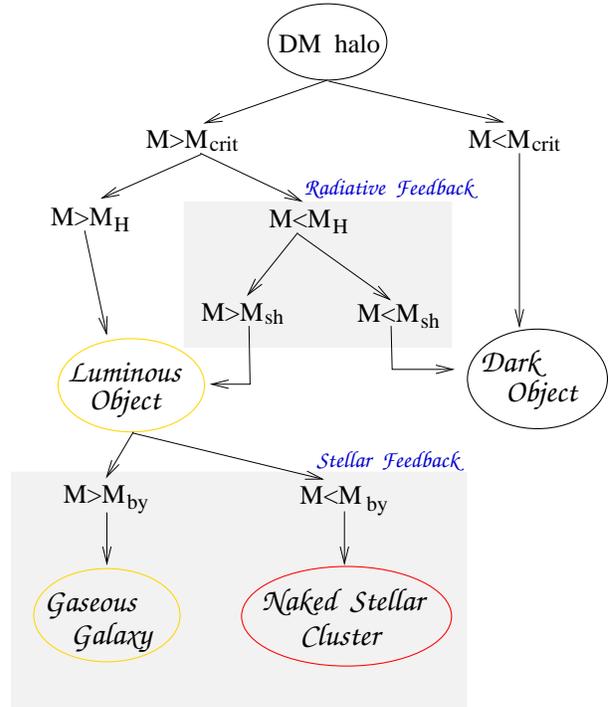, width=8cm}
  \end{center}
\caption{\label{fig3}{Possible evolutionary tracks of objects as
determined by the processes and feedbacks discussed in this paper}} 
\end{figure}

Fig. \ref{fig3} illustrates all possible evolutionary tracks and final fates
of primordial objects, together with the mass scales determined by the various
physical processes and feedbacks. We recall that there are four critical mass
scales in the problem: (i) $M_{crit}$, the minimum mass for an object to be
able to cool in a Hubble time; (ii) $M_H$, the critical mass for which hydrogen
Ly$\alpha$ line cooling is dominant;
(iii) $M_{sh}$, the characteristic mass
above which the object is self-shielded, and (iv)
$M_{by}$ the characteristic mass for stellar feedback, below which blowaway
can not be
avoided. Starting from a virialized dark matter halo, condition (i) produces
the first branching,  and objects failing to satisfy it will not collapse
and form only a negligible amount of stars. In the following,
we will refer to these objects as {\it dark objects}.
Protogalaxies with masses in the range $M_{crit} <
M < M_{H}$ are then subject to the effect of radiative feedback,
which could either impede the collapse of those of them with
mass $M<M_{sh}$, thus contributing to the
class of dark objects, or allow the collapse of the remaining ones ($M>M_{sh}$)
to join those with $M>M_H$ in the class of {\it luminous objects}. This is the
class of objects that convert a considerable fraction of their baryons in stars.
Stellar feedback causes the final bifurcation by inducing a blowaway of
the baryons
contained in luminous objects with mass $M<M_{by}$; this separates the class in
two subclasses, namely "normal" galaxies (although of masses comparable to
present day
dwarfs) that we dub {\it gaseous galaxies} and tiny stellar aggregates with
negligible traces (if any) of gas to which consequently we will refer to as
{\it naked stellar clusters}.

\section{Detecting PopIII Objects}

Before we proceed to show how the reionization of the universe is regulated
by stellar feedbacks, we briefly discuss possible strategies to 
observe PopIII objects.
Given their small mass, these objects are likely to be faint: for a reasonable
mass-to-light ratio for young galaxies $\approx 0.1$, their bolometric luminosity is
$\approx 2\times 10^{40}$~erg~s$^{-1}$. A Type II SN is typically
one hundred times brighter. Thus, for periods even longer than a year (taking
into account the time stretching $\propto (1+z)$ of the light curve)
the SNII outshines its host galaxy. Since Type II SNe         originate
from massive stars, a necessary requirement is that the (unknown) IMF of
these objects is flat enough to extend into this regime. Clearly, this
condition is satisfied in the local universe, but it is not necessarily
so in the conditions prevailing when the universe was young. Nevertheless, it
is intriguing to speculate about the observational perspectives to detect
very high $z$ SNe, a possibility to which our hopes to
investigate directly the primeval star/galaxy formation are closely tied.

\begin{figure}[ht]
  \begin{center}
    \epsfig{file=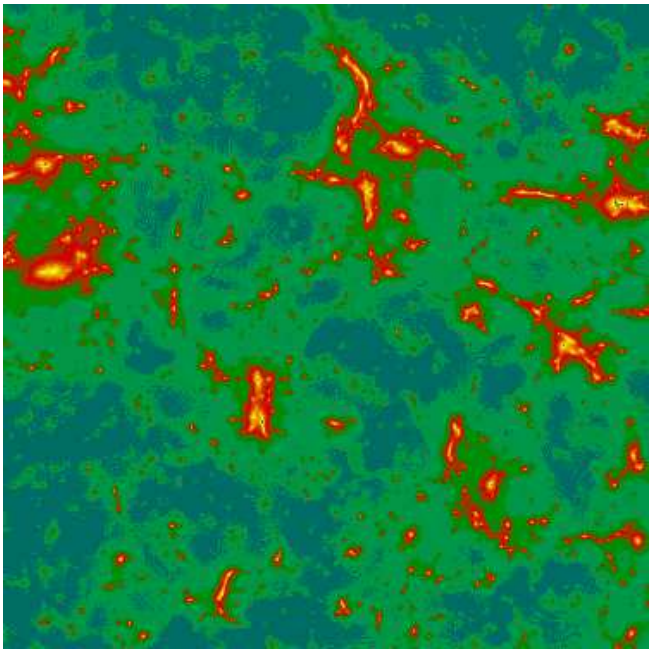, width=12cm}
  \end{center}
\caption{\label{fig4}{
Magnification map for a SCDM model
(4'$\times$ 4', corresponding to a NGST field) for a source
located at $z=10$. The magnification range is 0.7-50.
}}
\end{figure}

\begin{figure}[ht]
  \begin{center}
    \epsfig{file=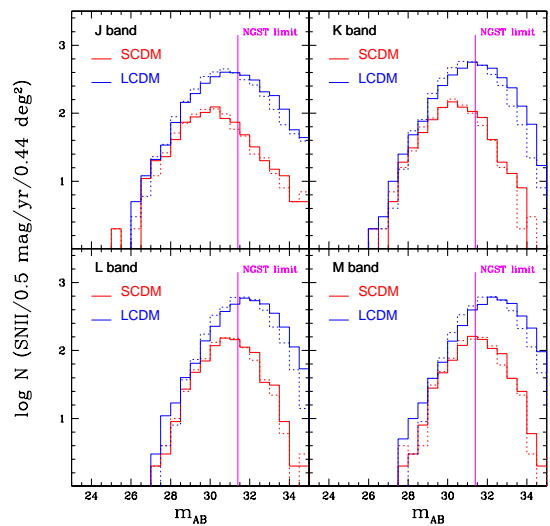, width=8cm}
  \end{center}
\caption{\label{fig5} Differential number counts for the two cosmological
models
considered ({\it thick solid line}: SCDM, {\it thin solid line}: LCDM)
as a function of apparent AB magnitude in $J$,$K$,$L$,$M$ bands.
Also shown is the NGST limiting magnitude.
{\it Dashed} curves neglect lensing magnification.}
\end{figure}

The advances in technology are making available a new generation of
instruments, some already at work and some in an advanced design phase,
which will dramatically increase our observational capabilities.
As representative of such class, we will focus on a particular instrument,
namely the Next Generation Space Telescope (NGST). In the following we
will try to quantify the expectations for the detection of high $z$ SNe.
In addition to what we can learn on the formation
of the first objects, Pop III SNII can provide crucial information on
cosmological
models due to their different predictions concerning the gravitational lensing
magnification patterns by the intervening matter in the universe.
The gravitational amplification might affect the SNII number counts and
should be therefore properly taken into account (Marri \& Ferrara 1998).
As an example, taken from the study of Marri, Ferrara \& Pozzetti (2000), 
we show the magnification map for the
SCDM model for a source located at redshift 10 in Fig. \ref{fig4}.
Fig. \ref{fig5} shows the differential SNII counts
[0.5~mag/ yr/ 0.44 deg$^2$] as a function of AB magnitude.
The four panels contain the curves for the SCDM and LCDM models and for
$J$, $K$, $L$, and $M$ bands, both including or neglecting
the effects of GL.
For comparison,
we plotted the NGST magnitude limit $AB=31.4$ (vertical line).
This is calculated by assuming a constant limiting flux
${\cal F}_{NGST}=10$~nJy in the wavelength range $1-5
\mu$m (\ie $J-M$ bands). This can be achieved, for a
8-m (10-m) mirror size and a S/N=5, in about $2.6\times 10^4$~s
($1.1\times 10^4$~s)\footnote{This result has been obtained using
the NGST Exposure Time Calculator, available at http://augusta.stsci.edu}.
Thus, NGST should be able to reach the peak of expected SNII count
distribution, which is located at $AB\approx 30-31$ for SCDM and
$AB\approx 31-32$ for LCDM (depending on the wavelength band).
The differences among the various bands are not particularly pronounced,
although  $J$ and $K$ bands present a larger number of luminous ($AB \simlt 27$)
sources, and therefore they might be more suitable for the experiment.
Furthermore, we point out that in the  $L$ and $M$ bands NGST,
with the current magnitude limit, will not be able to reach the peak
of expected SNII count distribution in the LCDM model.

\section{More feedback from dust formation}

In addition to the above mentioned feedback effects, 
SNII can also initiate molecular hydrogen formation on dust grain surfaces rather
than in the gas phase,
a more standard process in present day galactic environments.
At high redshift Type II SNe are the only possible sources of
dust, due to the short age of the universe and the long evolutionary
timescales characterizing more conventional dust sources, as for example
evolved  stars.
A detailed nucleation study of dust formation in the ejecta of
Type II SNe whose initial chemical composition is primordial has been very
recently carried on (Todini \& Ferrara, in preparation). This case is
suitable for the study of the very first SNII, as those originating in
PopIII objects. From this study we have been able to derive several
important properties characterizing the first solid particles produced
in the universe, as for example the total dust mass produced, the dust
composition and the grain size distribution. Preliminary results
show that in the range of SNII masses considered,
$M_{SN}=12-40 M_\odot$, typically 0.1 $M_\odot$ of dust is produced.
SNII towards the low end of the mass distribution mostly produce silicate
grains, whereas the most massive ones predominantly produce graphite grains;
this fact is easy to understand in terms of the C/O ratio of their ejecta.
The size distribution is approximated by a power-law for about two decades
in radius (from $\approx 1$\AA~ to 100\AA) and shows a smooth cutoff beyond
that grain radius. Thanks to these results we can now pose the following
question: what is the minimum amount of dust required in order for the
molecular hydrogen formation on grains to become competitive with the
gas phase one ? An order-of-magnitude answer can be obtained by comparing the
two formation rates. At the low densities relevant   here,
\HH   is formed in the gas phase mainly
via the channel $H+ e^- \rightarrow H^- +h\nu$, at rate $k_8$
(the rate coefficient $k_8$ is given in Abel \etal 1997);
formation via the \HH$^+$ channel, when
included, is found to be negligible in our case. Therefore the formation rate
in the gas phase is ${\cal R} \simeq k_8 n_{H^-} n_H$. The formation rate on
grain surfaces is instead given by ${\cal R}^d \simeq 0.5 \langle \gamma
c_s \sigma\rangle n_d n_H$, where $\gamma$ is the sticking coefficient,
$c_s$ is the sound speed in the gas, and $\sigma$ is the grain cross section.
The equality between the two rates can be cast into the following form:
${\cal D} = 0.1 \sqrt{T} x_e$, where ${\cal D}$ is the dust-to-gas ratio
normalized
to its Galactic value, $T$ is the gas temperature and $x_e$
the gas ionization fraction. For typical parameters of the PopIII objects,
\HH   production on dust grains becomes dominant once ${\cal D}$ is larger than
5\% of the local value. With the dust yields per SNII calculated above we then
conclude that only about 15 SNII are required to enrich in dust to this level
a primordial object. Clearly, early dust formation might play a role in
the formation of the first generation of objects.

\section{Reionization History Driven by Feedbacks}

In this final Section we summarize some of the results obtained by
CFGJ on the evolution of inhomogeneous reionization, regulated by
the feedback mechanisms introduced above and regulated by the molecular hydrogen
network. For a broader presentation the reader is advised to refer to CFGJ.
We have determined the spatial distribution
of the ionizing sources from high resolution numerical N-body simulations
within a periodic cube of comoving length $L=2.55h^{-1}$ Mpc for a
critical density cold dark matter model ($\Omega_0$=1, $h$=0.5 with
$\sigma_8$=0.6 at $z$=0).
This  allows us to describe the topology of the ionized and dissociated
regions at various cosmic epochs and derive the evolution of H, He, and \HH  
filling factors, soft UV background, cosmic star formation rate and the final
fate of ionizing objects. There are three free parameters in the computation:
$f_b$, the fraction of virialized baryons that is able to cool and
become available to form stars, $f_{\star}$ the star formation efficiency,
and $f_{esc}$ the photon escape fraction from the proto-galaxy.
The variation of these free parameters defines four different runs A-D,
whose parameter combination is shown in Tab. 1.
\begin{table}[hp]
\centerline{\begin{tabular}[t] {|l|l|l|l|l|l|l|r|} \hline
		  &             &             &           \\
			  {\em RUN} & {\em $f_b$} & {\em $f_{\star}$} & {\em $f_{esc}$} \\
		  &             &                   &           \\ \hline
		  A  & 0.08 & 0.15 & 0.2 \\
		  B  & 0.08 & 0.05 & 0.2 \\
		  C  & 1.00 & 0.15 & 0.2 \\
	      D  & 0.08 & 0.15 & 0.1 \\ \hline
\end{tabular}}
\caption{Parameters of the calculation: fraction of virialized baryons
that are able to cool and be available for star formation, $f_b$; star
formation efficiency, $f_{\star}$; photon escape fraction, $f_{esc}$.}
\label{tab1}
\end{table}

\subsection{Filling factors}

\begin{figure}[ht]
  \begin{center}
    \epsfig{file=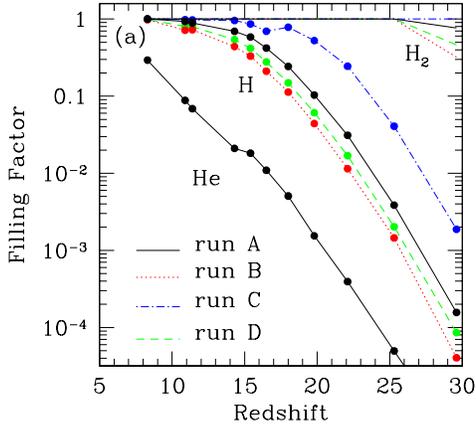, width=8cm}
  \end{center}
\caption{\label{fig6}{Evolution of the  dissociated molecular hydrogen (upper set of lines),
ionized atomic hydrogen (middle set) and doubly ionized helium
(bottom line)  filling factor as a function of redshift for different runs:
A (solid line), B (dotted), C (dashed-dotted) and D (dashed).
(b) same as (a) for runs: A (solid line), A1 (dotted)
and A2 (dashed-dotted).}}
\end{figure}

The filling factors of the dissociated H$_2$ and ionized H and He are defined as
the box volume fraction occupied by those species.
The results are shown in Fig. \ref{fig6}, for the different
runs. The intergalactic relic molecular hydrogen is found to be
completely dissociated at very high redshift ($z\approx 25$) independently
of the parameters of the simulation. This descends from the fact that
dissociation spheres are relatively large and overlap at early times.
Ionization spheres are instead always smaller than dissociation ones
and complete reionization occurs considerably later.
Except for run C, when reionization occurs by $z$$\approx$15,
primordial galaxies are able to reionize the IGM at a redshift
$z$$\approx$10.
The filling factor is approximately linearly
proportional to the number of sources but has a stronger (cubic) dependence
on the ionization sphere radius. Thus, although the number
of (relatively small) luminous objects present increases with redshift,
the filling factor is only boosted by the appearance of larger, and therefore
more luminous, objects which can ionize more efficiently.
The subsequent flattening at lower redshifts, is obviously due to the
fact that when the volume fraction occupied by the ionized gas becomes
close to unity most photons are preferentially used to sustain the reached
ionization level rather than to create new ionized regions.
The reionization is basically driven by objects collapsed through
H line cooling, while small mass PopIII stars   play only a
minor role and even in the absence of a radiative negative feedback
they would not be able to  reionize the IGM.
Note that only about 2\% of the stars
observed at $z=0$ is required to reionize the universe completely. This
corresponds to an
average IGM metallicity (assuming a Salpeter IMF) at redshift $z$$\approx$10
equal
to $\langle Z \rangle \approx 3 \times 10^{-4} Z_\odot$.

\subsection{What is the fate of (re)ionizing sources ?}
\begin{figure}[ht]
  \begin{center}
    \epsfig{file=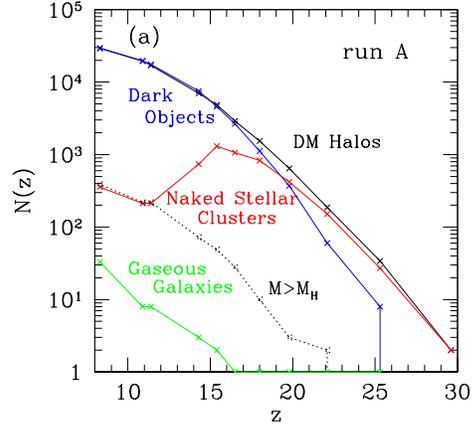, width=8cm}
  \end{center}
\caption{\label{fig7}{Number evolution of different objects in the
simulation box for run A.}}
\end{figure}

The last point we make here concerns the final fate of the objects responsible
for the reionization of the universe in terms of the evolutionary tracks
introduced above; this discussion is summarized in Fig. \ref{fig7}.   
The straight lines represent, from the top to the bottom,
the number of dark matter halos, dark objects, naked
stellar clusters and gaseous galaxies, respectively.
The dotted curve represents the
number of luminous objects with large enough mass
($M>M_H$) to make the H line cooling efficient and become insensitive to
the negative feedback. We remind that the naked stellar
clusters are the luminous objects with $M<M_{by}$, while the
gaseous galaxies are the ones with $M>M_{by}$; thus, the number of
luminous objects present at a certain redshift is given by the sum of
naked stellar clusters and gaseous galaxies.
We first notice that the majority of
the luminous objects that are able to form at high redshift will
experience blowaway, becoming naked stellar clusters, while only a minor
fraction, and only at $z\simlt$15, when larger objects start to form,
will survive and become gaseous galaxies.
An always increasing number of luminous objects is forming with
decreasing redshift, until $z$$\approx$15, where a flattening is seen. This
is  due to the fact that the dark matter halo mass function is
still dominated by small mass objects, but a large fraction of them cannot
form due to the following combined effects: i) towards lower
redshift
the critical mass for the collapse ($M_{crit}$) increases and fewer
objects satisfy the condition $M>M_{crit}$; ii) the radiative feedback
due to either the direct dissociating flux or the SUVB 
increases at low redshift as the SUVB intensity reaches values
significant for the negative feedback effect.
When the number of luminous objects becomes
dominated by objects with $M>M_H$, by
$z$$\approx$10 the population of luminous objects grows again, basically because
their
formation is now unaffected by negative feedback.
A steadily increasing number   of objects is prevented from forming stars
and remains dark; this population is about $\approx$99\% of the total
population of dark matter halos at $z$$\approx$8. This is also due to the
combined effect of points i) and ii) mentioned above.
This population of halos which have failed to produce
stars could be identified with the low mass
tail distribution of the dark galaxies that reveal
their presence through gravitational lensing of quasars.
It has been argued in the recent literature that this
population of dark galaxies outnumbers normal galaxies by a substantial
amount, and   Fig. \ref{fig7} supports this view.

\section{Summary}

We have discussed several feedback effects, mostly related to   the presence
of massive stars in early formed objects, and analyzed their impact on 
galaxy formation and reionization of the universe. A brief summary of the
findings of this paper (for a more detailed presentation see Ciardi \etal 2000)
follows below:

{$\bullet$} The first objects affect subsequent galaxy formation both via
{\it radiative} and {\it stellar} feedbacks due to
ionizing/ \HH--photodissociating radiation and mechanical energy input due
to massive stars and supernovae, respectively

{$\bullet$} Molecular hydrogen is photodissociated but reformed behind SN shocks
during blowaway/blowout events; dust can be formed in SNII ejecta, allowing
for \HH formation on grain surfaces rather than in the gas phase.

{$\bullet$} PopIII objects could be detected via their SNII with NGST;
Up to $z=15$, the (SCDM, LCDM) models
predict a total number of (857, 3656) SNII/yr in 100 surveyed
$4'\times 4'$ fields of the {\it Next Generation Space Telescope}.

{$\bullet$} Galaxies are able to reionize the neutral atomic hydrogen by
a redshift $z$$\approx$10, while molecular hydrogen is completely
dissociated at very high redshift ($z$$\approx$25).

{$\bullet$} IGM reionization is basically driven by objects collapsed through
H line cooling ($M>M_H$), while small mass objects ($M<M_H$) play only a
minor role and even in the absence of a radiative negative feedback
they would not be able the reionize the IGM.

{$\bullet$} A consistent         fraction of halos   is prevented from forming stars
by either the condition $M<M_{crit}$ or the effect of radiative feedback;
this population of dark objects reaches $\approx$99 \% of the
dark matter halo population at $z$$\approx$8.

\begin{acknowledgements}
We would like to thank our collaborators T. Abel, J. Dove, F. Governato, A. Jenkins,
M.-M. MacLow, L. Pozzetti, Y. Shchekinov and M. Shull for sharing their enthusiasm 
with us.
\end{acknowledgements}

\end{document}